\documentclass[12pt,preprint]{aastex}      

\usepackage[latin1]{inputenc}
\usepackage{lscape}

\shorttitle{Multiplicity, Disks and Jets in NGC~2071}
\shortauthors{Carrasco-Gonz\'alez et al.}

\begin{document}

\title{Multiplicity, Disks, and Jets in the NGC 2071 Star-Forming Region}

\author{
Carlos~Carrasco-Gonz\'alez\altaffilmark{1},
Mayra~Osorio\altaffilmark{2},
Guillem~Anglada\altaffilmark{2},
Paola~D'Alessio\altaffilmark{3},
Luis~F.~Rodr\'{\i}guez\altaffilmark{3},
Jos\'e~F.~G\'omez\altaffilmark{2,4},
Jos\'e~M.~Torrelles\altaffilmark{5}
}

\altaffiltext{1}{Max-Planck-Institut f\"ur Radioastronomie (MPIfR), Auf dem H\"ugel 69, 53121, Bonn, Germany; carrasco@mpifr-bonn.mpg.de}

\altaffiltext{2}{Instituto de Astrof\'{\i}sica de Andaluc\'{\i}a, CSIC, Camino Bajo de Hu\'etor 50, E-18008 Granada, Spain}

\altaffiltext{3}{Centro de Radioastronom\'{\i}a y Astrof\'{\i}sica UNAM, Apartado Postal 3-72 (Xangari), 58089 Morelia, Michoac\'an, M\'exico}

\altaffiltext{4}{On sabbatical leave at CSIRO Astronomy \& Space Science, Marsfield, NSW 2122, Australia}

\altaffiltext{5}{Instituto de Ciencias del Espacio (CSIC)-UB/IEEC, Universitat de Barcelona, Mart\'{\i} i Franqu\`es 1, E-08028 Barcelona, Spain}

\begin{abstract}

We present centimeter and millimeter observations of the NGC 2071 star-forming region performed with the VLA and CARMA. We detected counterparts at 3.6 cm and 3 mm for the previously known sources IRS 1, IRS 2, IRS 3, and VLA 1. All these sources show SEDs dominated by free-free thermal emission at cm wavelengths, and thermal dust emission at mm wavelengths, suggesting that all of them are associated with YSOs. IRS 1 shows a complex morphology at 3.6 cm, with changes in the direction of its elongation. We discuss two possible explanations to this morphology: the result of changes in the direction of a jet due to interactions with a dense ambient medium, or that we are actually observing the superposition of two jets arising from two components of a binary system. Higher angular resolution observations at 1.3 cm support the second possibility, since a double source is inferred at this wavelength. IRS 3 shows a clear jet-like morphology at 3.6 cm. Over a time-span of four years, we observed changes in the morphology of this source that we interpret as due to ejection of ionized material in a jet. The emission at 3 mm of IRS 3 is angularly resolved, with a deconvolved size (FWHM) of $\sim$120 AU, and seems to be tracing a dusty circumstellar disk perpendicular to the radio jet. An irradiated accretion disk model around an intermediate-mass YSO can account for the observed SED and spatial intensity profile at 3 mm, supporting this interpretation.

 \end{abstract}


\section{Introduction}

 NGC~2071 is a reflection nebula located at a distance of 390~pc in the L1630 molecular cloud of Orion B (Anthony-Twarog 1982). A powerful molecular outflow was observed approximately 4$\arcmin$ north of the NGC~2071 reflection nebula. This outflow is oriented in the NE-SW direction and extends $\sim$15$\arcmin$ in length and $\sim$120~km~s$^{-1}$ in velocity. It has been extensively studied in CO (Bally 1982; Snell et al. 1984; Scoville et al. 1986; Moriarty-Schieven 1989; Margulis \& Snell 1989; Kitamura et al. 1990; Chernin \& Masson 1992; Chernin \& Welch 1995; Houde et al. 2001; Stojimirovic et al. 2008), HI (Bally\& Stark 1983), OH (Ruiz et al. 1992), and SiO and CH$_3$OH (Garay et al. 2000). Moreover, shock-excited molecular hydrogen emission at 2.12~$\mu$m shows a spatial extent similar to the CO outflow, as well as several additional outflows in the region (Eisl{\"o}ffel 2000). At the center of the outflow there is a $\sim$30$\arcsec$ diameter infrared cluster (Persson et al. 1981) with a total luminosity of 520~L$_\sun$ (Butner et al. 1990), which has been identified as an intermediate-mass star-forming region. Continuum emission at 6 cm wavelength has been detected associated with three of the IR sources (IRS 1, 2, and 3; Snell \& Bally 1986). Sources IRS~1 and IRS~3 are also associated with 1.3~cm continuum and water maser emission (Torrelles et al. 1998). In IRS~1, both the 1.3 cm continuum and the water maser emission seem to be tracing a radio jet. Interestingly, in IRS~3, while the 1.3~cm emission clearly traces a radio jet, the water maser spots seem to be tracing a perpendicular compact disk with a radius of $\sim$20~AU (Torrelles et al. 1998). Recently, Trinidad et al. (2009) also observed the region at 1.3 cm and found similar results for IRS 1 and IRS 3. In the observations by Trinidad et al. (2009), a new 1.3 cm compact source with no previous known IR counterpart was detected between IRS 1 and IRS 3, which they named as VLA 1.
 
 In this paper we present new high angular resolution observations at cm and mm wavelengths towards the star-forming region in NGC~2071 to disentangle the nature of the different young stellar objects (YSOs) in the cluster. We used the Very Large Array (VLA) and the Combined Array for Research in Millimeter-wave Astronomy (CARMA) in their most extended configurations. 

\section{Observations}

\subsection{VLA Observations}

 Continuum observations at 3.6 cm were made with the VLA of the National Radio Astronomy Observatory (NRAO)\footnote{The NRAO is a facility of the National Science Foundation operated under cooperative agreement by  Associated Universities, Inc.}\ using the A configuration and a effective total bandwidth of 200~MHz. The observations were carried out in five epochs, ranging from 1995 August 12 to 1999 July 3. In Table \ref{Tab1} we show a summary of the observation parameters used in each epoch. The same phase center ($\alpha$(J2000)=05$^h$47$^m$04.784$^s$, $\delta$(J2000)=00$^\circ$21$\arcmin$43.103$\arcsec$) and phase calibrator (J0541$-$056) were used for all the epochs. Flux calibration was achieved by observing 3C286 and 3C48. Data editing and calibration were carried out using the AIPS package of the NRAO, following the standard VLA calibration procedures. Resolved models for the flux calibrators were used. After a first calibration, data from each epoch were self-calibrated in amplitude and phase. After calibration, we made maps for each epoch with natural and uniform weightings. We also made natural and uniform weighting maps from the data after concatenation of all the 3.6 cm epochs. 

 Additional A configuration data at 20, 6 and 2 cm were taken from the VLA archive (see Table \ref{Tab1}) and calibrated following standard VLA procedures. 
\subsection{CARMA Observations}

 Continuum data at 3~mm were taken using the CARMA interferometer\footnote{Support for CARMA construction was derived from the states of California, Illinois, and Maryland, the James S. McDonnell Foundation, the Gordon and Betty Moore Foundation, the Kenneth T. and Eileen L. Norris Foundation, the University of Chicago, the Associates of the California Institute of Technology, and the National Science Foundation. Ongoing CARMA development and operations are supported by the National Science Foundation under a cooperative agreement, and by the CARMA partner universities.} (Mundy \& Scott 2000). We obtained B and A configuration data which provide angular resolutions of $\sim$0$\farcs$5 and $\sim$0$\farcs$3, respectively. Continuum data were recorded in six windows of $\sim$500 MHz width, providing a combined frequency coverage of 99.75-100.67 GHz and 102.90-103.82 GHz for the lower and upper sidebands of the receiver, respectively. Each band consists of 15 channels. Flux and bandpass calibration were achieved by observing Uranus and J0530+135, respectively. Data editing and calibration were performed using the MIRIAD package (Sault et al. 1995). A summary of the observational parameters is given in Table \ref{Tab1}.

\section{Results and Discussion}

\begin{figure}[!h] 
\begin{center}

\includegraphics[width=\textwidth]{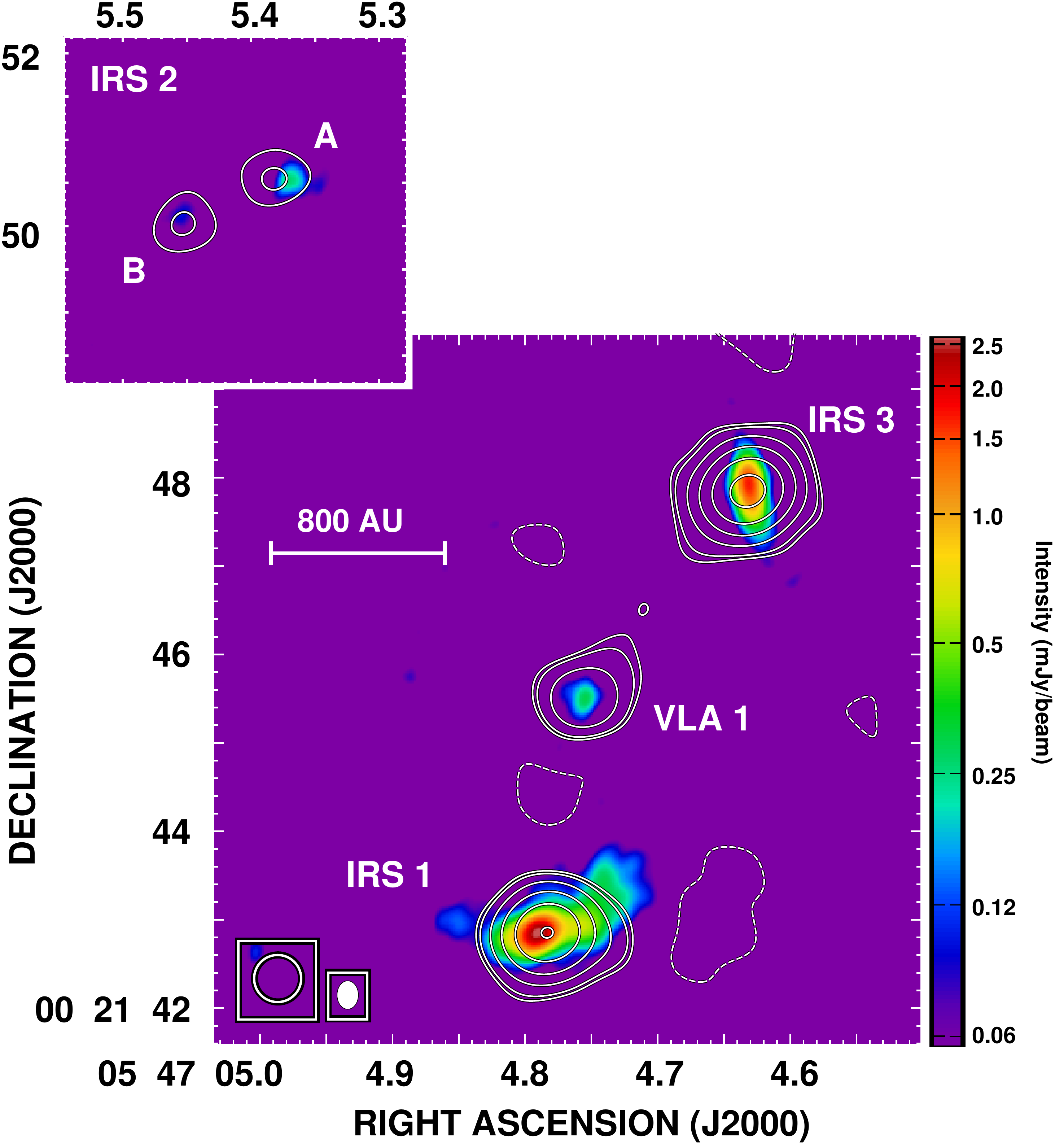}  

\caption{Superposition of the B configuration CARMA 3 mm continuum emission map (contours) over the A configuration VLA 3.6 cm continuum emission map (colors) of the NGC~2071 region. The 3.6 cm map was made by concatenating the uv data from the five epochs. The rms of the 3.6~cm map is 19~$\mu$Jy~beam$^{-1}$. Contours are $-$3, 3, 4, 8, 16, 32, and 64 times the rms of the 3~mm map, 370 $\mu$Jy~beam$^{-1}$. Both maps were made using natural weighting. Beam sizes are indicated in the lower left corner, with the filled ellipse showing that of the VLA data.}

\label{Fig1}  
\end{center}
\end{figure}

\begin{figure}[!h]
\begin{center}

\includegraphics[width=\textwidth]{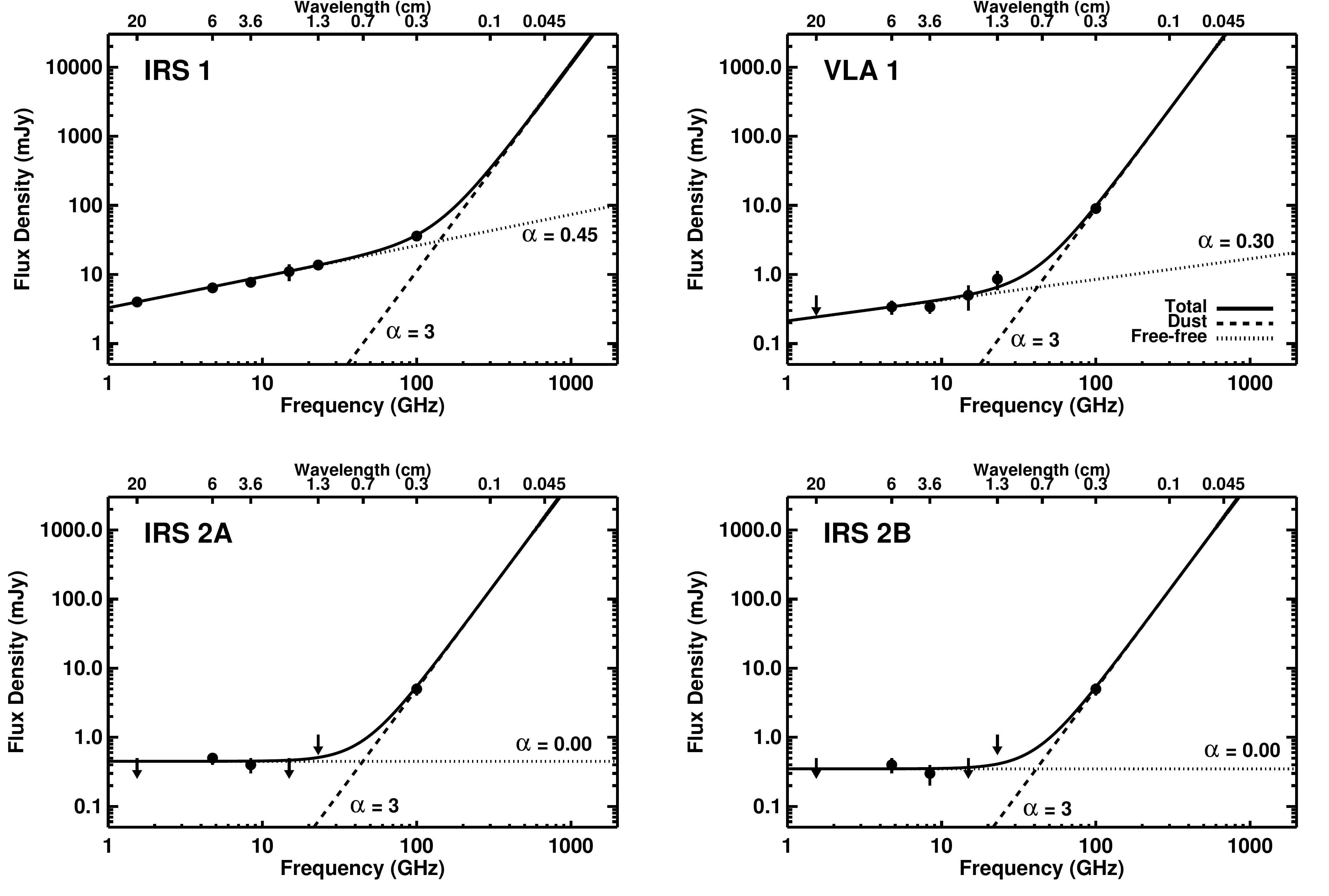}  

\caption{Spectral energy distributions of sources IRS 1, VLA 1, IRS 2A, and IRS 2B from 20~cm to 3~mm. The data points have been fitted as the sum of two power laws, one of them with a fixed slope of $+$3 (dust).} 

\label{Fig2}  
\end{center}
\end{figure}

 In Figure \ref{Fig1} we show a superposition of the CARMA 3 mm continuum map (contours) over the VLA 3.6 cm continuum map (colors) of the NGC~2071 region. As can be seen in this figure, we detect counterparts of the sources IRS~1, IRS~2 (this source is observed to have two components named as A and B), IRS~3 and VLA~1 at both wavelengths. In Table \ref{Tab2} we give the flux densities of the sources in the range from 20 cm to 3 mm. In Figure \ref{Fig2} we show the spectral energy distribution (SED) in the cm-mm wavelength range obtained from our data, for sources IRS~1, VLA~1, IRS~2A, and IRS~2B (the SED of IRS~3 will be discussed in section \ref{SectionIRS3}). All the sources show the typical radio spectra found in YSOs consistent with free-free emission at cm wavelengths plus a thermal dust contribution at mm wavelengths (e.g., Rodr\'{\i}guez 1995, Anglada 1996). In the following subsections, we discuss in more detail the results obtained for each of these sources.
 
\subsection{IRS 2: A binary system of YSOs} \label{SectionIRS2}

 Source IRS~2 was previously detected at 6~cm as a single source by Snell \& Bally (1986) (C configuration data). In our higher angular resolution maps (A configuration data), this source is resolved into two weak components (A and B) separated by 1$\farcs$3 ($\sim$500~AU). Both components appear compact and show emission at 6~cm, 3.6~cm and 3~mm (see Figure \ref{Fig1} and Table \ref{Tab2}). The flux densities and upper limits in the 20 cm to 1.3 cm wavelength range are consistent with flat spectral indices ($\alpha\simeq$0; S$_\nu$~$\propto$~$\nu^\alpha$) for both sources, which suggest optically-thin, free-free thermal emission. The high flux densities at 3 mm indicate that the emission at mm wavelengths is dominated by dust thermal emission. We suggest that these two sources form a binary system of YSOs. 
   
\subsection{VLA 1: A very young, deeply embedded YSO} \label{SectionVLA1}

 Source VLA~1 has not been previously detected at IR wavelengths, but it was detected at 1.3~cm by Trinidad et al. (2009). These authors also detected water maser emission associated with this source. We detected counterparts for VLA 1 at all wavelengths between 3~mm and 6~cm (see Figure \ref{Fig1} and Table \ref{Tab2}). The SED of this source also indicates free-free emission in the 20~cm to 1.3~cm wavelength range ($\alpha\simeq$0.3) and thermal dust emission at mm wavelengths. Therefore, this source is most probably another YSO associated with the region. 
 
 Source VLA 1 is also associated with X-ray emission with the spectrum of a heavily obscured object (Skinner et al 2009). This result and the non-detection of this source at IR wavelengths, suggest that VLA 1 traces a younger, more embedded protostar than the other radio sources in the region.
 
\subsection{IRS 1: A binary jet?} \label{SectionIRS1}

 IRS 1 has been classified as a class I source (Skinner et al. 2009). It was found to be associated with 1.3~cm continuum emission and water maser emission by Torrelles et al. (1998). These authors describe the morphology of the continuum emission in IRS~1 as consisting of a main body elongated east-west and a weak protuberance of emission to the north-east. Additionally, the water maser spots associated with this source appear spread along a direction close to the direction of the elongation of the continuum emission. Therefore, Torrelles et al. (1998) proposed that both the continuum emission and the water maser spots are tracing a radio jet. 

 The flux densities in the 20~cm to 1.3~cm wavelength range obtained from our observations are consistent with a spectral index of $\sim$0.45, which is suggestive of free-free emission from an ionized jet. At 3~mm, the source appears unresolved even in the uniform weighting map, and with a flux density $\sim$30\% higher than expected from the extrapolation of the free-free emission. This is consistent with a fraction of the emission at 3~mm arising from thermal dust emission from a dusty disk or envelope.

\begin{figure}[!h] 
\begin{center}

\includegraphics[height=0.8\textheight]{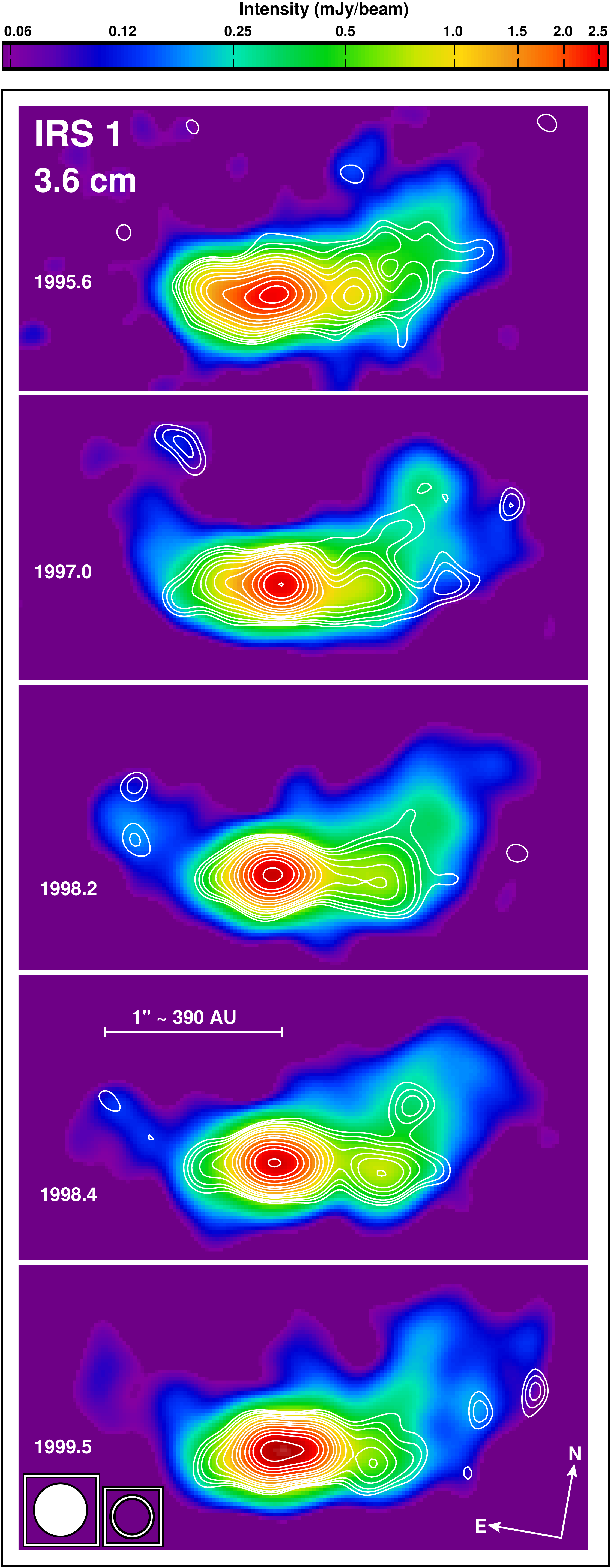}  

\caption{VLA 3.6 cm maps of IRS 1 at several epochs. For each epoch, we show the superposition of the uniform weighting map (contours) over the natural weighting map (colors). The maps of the different epochs were reconstructed with the same circular beams, 0$\farcs$30 (natural weighting maps) and 0$\farcs$22 (uniform weighting maps). Contours are 4, 5, 6, 8, 10, 12, 16, 20, 24, 32, 40, 48, 64, 80, and 96 times 36~$\mu$Jy~beam$^{-1}$.}

\label{Fig3}  
\end{center}
\end{figure}

\begin{figure}[!h]  
\begin{center}

\includegraphics[width=0.9\textwidth]{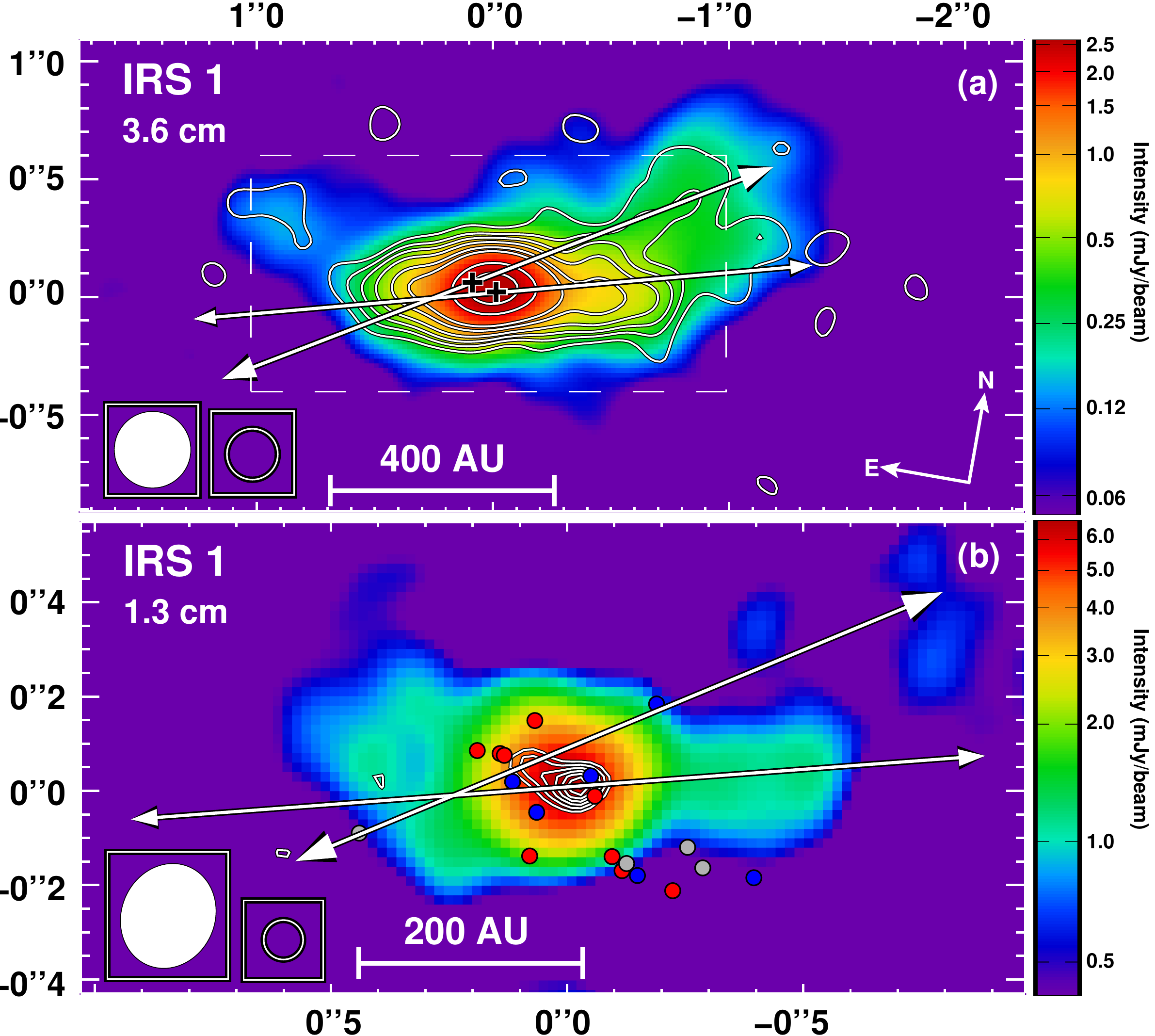}  

\caption{VLA maps of IRS 1. \textbf{(a)} Superposition of the 3.6~cm uniform weighting map (contours; synthesized 
beam=0$\farcs$22$\times$0$\farcs$22) over the 3.6~cm natural weighting map (colors; synthesized beam=0$\farcs$30$\times$0$\farcs$30). These maps were made after concatenation of data from all the observed epochs. The rms of the natural weighting map is 19~$\mu$Jy~beam$^{-1}$. Contours are $-$3, 3, 5, 7, 10, 15, 20, 40, and 60 times the rms of the uniform weighting map, 30 $\mu$Jy~beam$^{-1}$. Crosses mark the positions of the proposed protobinary system suggested by the 1.3~cm observations shown in panel (b). Solid lines mark a possible orientation of the two proposed jets (see section \ref{SectionIRS1}). \textbf{(b)} Superposition of the VLA 1.3 cm uniform weighting map (contours; synthesized beam=0$\farcs$09$\times$0$\farcs$09) over the 1.3~cm natural weighting map made with a tapering of 1000 k$\lambda$ (colors; synthesized beam=0$\farcs$22$\times$0$\farcs$20, P.A.=$-$13$^\circ$). The rms of the map made with tapering is 167~$\mu$Jy~beam$^{-1}$. Contours are $-$3, 3, 4, 5, 6, 7, 8, and 9 times the rms of the uniform weighting map, 400~$\mu$Jy~beam$^{-1}$. Water masers spots (Torrelles et al. 1998) are marked with red (V$_{LSR}$ $>$ 11.8 km s$^{-1}$), grey (7.8 km s$^{-1}$ $<$ V$_{LSR}$ $<$ 11.8 km s$^{-1}$) and blue (V$_{LSR}$ $<$ 7.8~km~s$^{-1}$) circles. In both panels, the (0,0) position corresponds to $\alpha$(J2000)=05$^h$47$^m$04.791$^s$, $\delta$(J2000)=00$^\circ$21$\arcmin$42.81$\arcsec$.}

\label{Fig4}  
\end{center}
\end{figure}

 In Figure \ref{Fig3} we show the 3.6~cm VLA maps of IRS~1 obtained in the five epochs. The source is elongated in a E-W direction, suggesting that it traces a radio jet in that direction. However, in the natural weighting map (color maps in Figure \ref{Fig3}), the emission at both edges of the radio jet bends to the north. This complex morphology is better seen in our highest sensitivity 3.6~cm maps of source IRS~1 (maps made by concatenating data from all the epochs) shown in Figure \ref{Fig4}a. In the uniform weighting map, which has higher angular resolution, the emission seems to be divided in two different structures: the core of the emission seems to follow an E-W direction, while there is some emission that traces a sort of filament in a SE-NW direction (see Figure \ref{Fig4}a). 
 
 In order to explain this morphology, we have analyzed two possibilities. A first possibility is that a single YSO drives a jet oriented in the E-W direction. At a distance of $\sim$300~AU from the driving source, both lobes of the jet strongly interact with dense material in the ambient medium. These shocks could result in a change of the direction of both lobes of the jet resulting in the curved morphology seen in our VLA maps. Since it is required that both lobes of the jet interact with dense gas, this implies that IRS~1 should be completely surrounded by a high density structure, with IRS~1 being embedded in a sort of cavity. Unfortunately, the presence of such a high density structure could not be tested with the present data. High angular resolution observations with a high-density molecular gas tracer would be necessary in order to test its presence. 
 
 A second possibility is that the emission from IRS~1 is actually arising from a binary jet, i.e., two jets, each of them driven by a different component in a close binary system. We note that the morphology of the IRS~1 source at cm wavelengths resembles that of the L1551-IRS~5 source (see Figure 1 of Rodr\'{\i}guez et al. 2003), which is known to be a binary jet driven by a binary system with a separation of $\sim$45~AU (Rodr\'{\i}guez et al. 1998, 2003; Lim \& Takakuwa 2006; Pyo et al. 2009). If IRS~1 in the NGC~2071 region is a similar case, then the binary system would remain unresolved with the angular resolution of the 3.6~cm observations. A way to support this possibility would be to resolve the core of the emission into a double source. To this aim, we reanalyzed the 1.3~cm data of Torrelles et al. (1998). In Figure \ref{Fig4}b we show (in colors) a map made with a tapering of 1000~k$\lambda$, which is more sensitive to weak extended emission than the maps shown in Torrelles et al. (1998). This map shows essentially the same morphology than the map at 3.6~cm (except for a 
slight change in the size of the source due to opacity effects). In the same figure, we show, superimposed in contours, a 1.3~cm map made with uniform weighting, which has the highest angular resolution attainable with the data at this wavelength (beam$\simeq$0$\farcs$09; we used uniform weighting, while the map shown in Torrelles et al. 1998 was obtained with ROBUST=0 in AIPS). As can be seen in the figure, this uniform weighting map shows emission from a compact source with an elongation to the NE. This elongation could be interpreted as due to emission from a second component, resulting in a binary system with a separation of $\sim$0$\farcs$11 ($\sim$40~AU) that is marginally resolved by the 0$\farcs$09 beam of the map. Note that this separation is also very similar to that found in the L1551-IRS~5 binary system. These results would support the possibility of a binary jet also in the IRS~1 source. In Figure \ref{Fig4} we show a possible configuration for the orientation of the two jets that approximately follows the cm emission. However, all these results should be confirmed with observations with higher angular resolution and sensitivity.
 
\subsection{IRS 3: A disk/jet system} \label{SectionIRS3}

 Source IRS~3 was previously detected by Torrelles et al. (1998) as a 1.3~cm continuum source showing a clear jet-like morphology. These authors also detected a group of water maser spots associated with IRS~3. The maser spots are distributed in a strip of 0$\farcs$1 size ($\sim$40~AU) centered on the continuum source, and almost perpendicular to its major axis. In addition, the LSR velocity of the masers presents a clear gradient along the major axis of their spatial distribution. From these results, Torrelles et al. proposed that while the 1.3~cm continuum emission from IRS~3 is tracing a radio jet, the water maser spots are tracing a rotating circumstellar disk with a radius of $\sim$20~AU which is seen almost edge-on. The orientation of the cm emission of IRS~3 (roughly NE-SW) is similar to that of the most prominent H$_2$ jet detected in the region, and therefore, this source is most likely driving this large scale jet (Eisl\"offel 2000). 
  
\begin{figure}[!h]  
\begin{center}

\includegraphics[width=\textwidth]{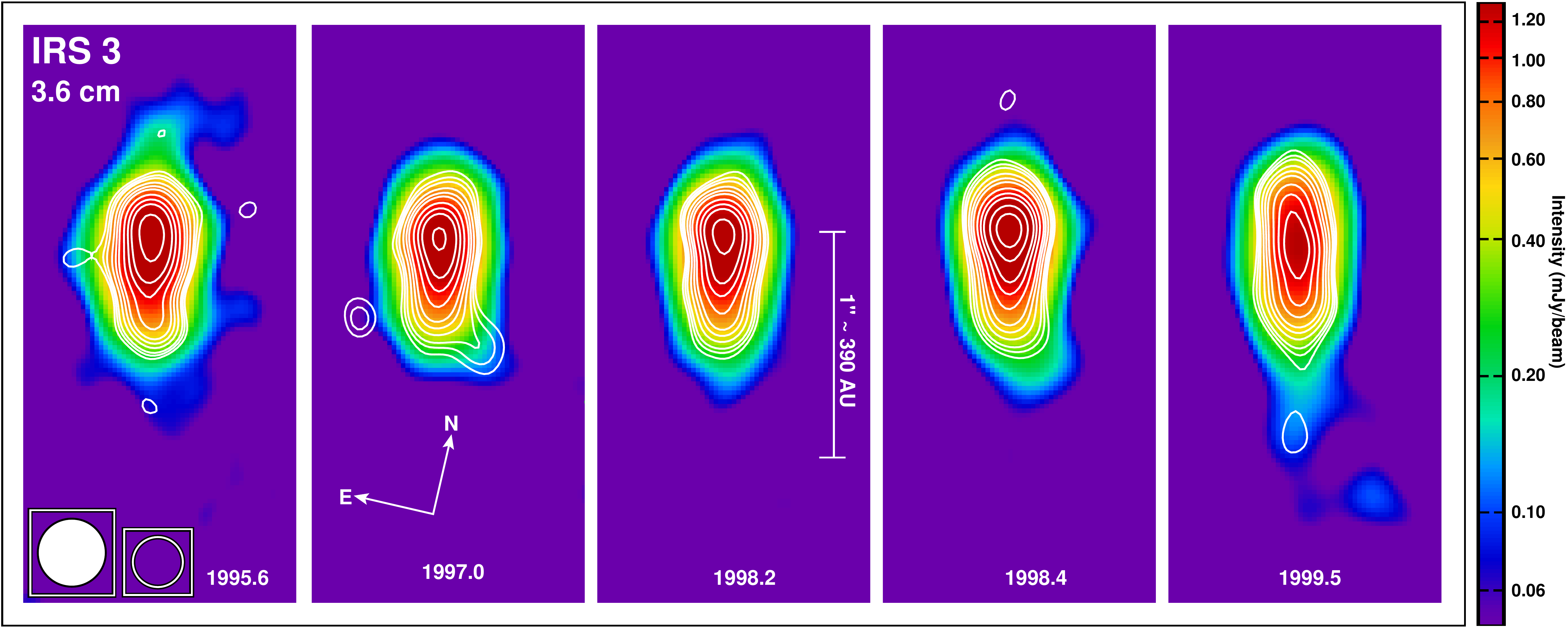}  

\caption{VLA 3.6 cm maps of IRS 3 at several epochs. For each epoch, we show the superposition of the uniform weighting map (contours) over the natural weighting map (colors). The maps of the different epochs were reconstructed with the same circular beams, 0$\farcs$30 (natural weighting maps) and 0$\farcs$22 (uniform weighting maps). Contours are 4, 5, 6, 8, 10, 12, 16, 20, 24, 32, 40, 48, 64, 80, and 96 times 36~$\mu$Jy~beam$^{-1}$.}
\label{Fig5}  
\end{center}
\end{figure}

\begin{figure}[!h]  
\begin{center}

\includegraphics[height=0.8\textheight]{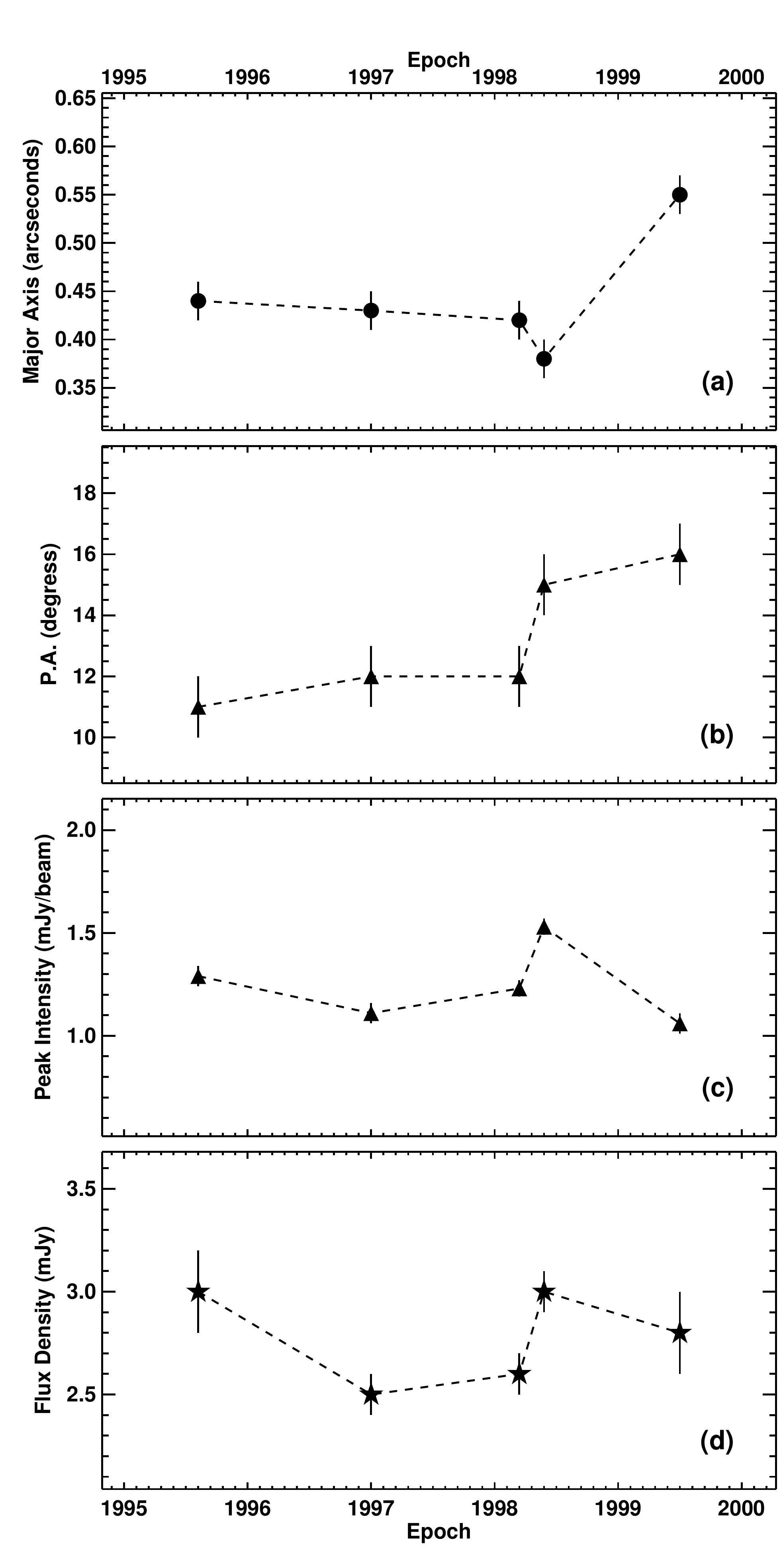}  

\caption{Monitoring of the deconvolved size of the major axis (panel a), position angle (panel b), peak intensity (panel c) and flux density (panel d) of IRS 3 at 3.6 cm. }
\label{Fig6}  
\end{center}
\end{figure}

 In Figure \ref{Fig5} we show VLA maps at 3.6~cm of IRS~3 obtained in the five epochs. In Table \ref{Tab3} and Figure \ref{Fig6} we show the deconvolved size of the major axis, the P.A., the peak intensity and the flux density of IRS~3 obtained from Gaussian fits to the uniform weight maps of each epoch. It can be seen than in the 1998.4 epoch, there is an increase of the flux density and the peak intensity of the source. Also in this year, there is a small change in the P.A. of the source, from $\sim$12$^\circ$ to $\sim$15$^\circ$. Finally, in the last epoch (1999.5) there is an increase in the size of the source. We interpret this behavior as due to an ejection of ionized material together with precession of the jet. The ejection of material would have taken place around 1998.3, since it was in this epoch that the peak intensity and the flux density increased. As the ejected material moves away from the central object, the size of the radio source would increase, in agreement with what is observed in the last epoch (1999.5). From the difference in size between 1998.4 and 1999.5, we estimated a velocity of $\sim$140~km~s$^{-1}$, which is of the same order than the velocities found in other jets from protostars. The coincidence of the ejection with the change in the P.A. of the jet (that persisted until the last epoch of observation), suggests that this behavior could be due to a (small) precession of the jet axis. This is consistent to what is observed in the large scale H$_2$ jet, which also shows hints of precession (see Figure 3 of Eisl\"offel 2000). 
    
\begin{figure}[!h]
\begin{center}

\includegraphics[width=\textwidth]{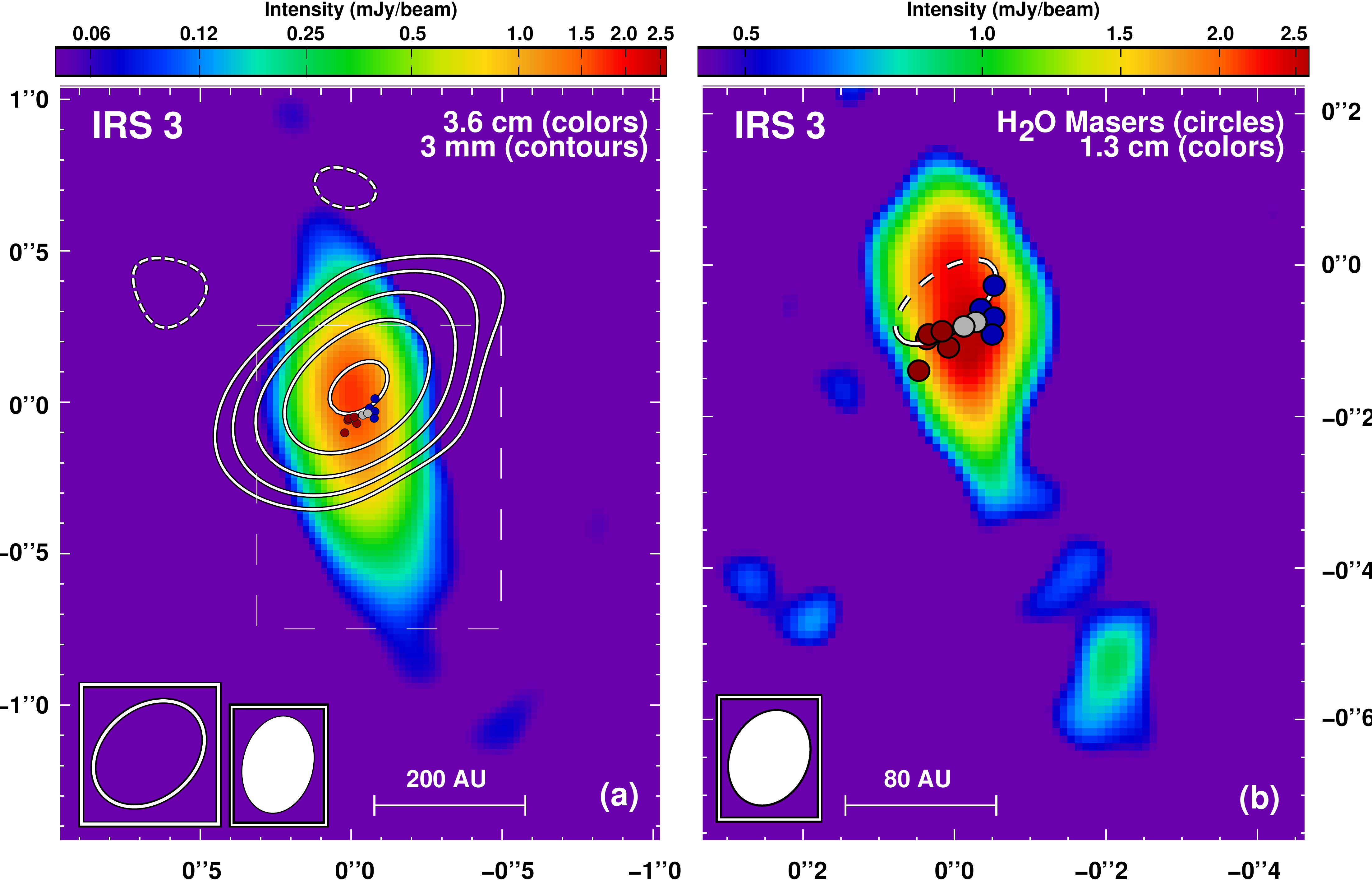}  

\caption{\textbf{(a)} Superposition of the 3 mm CARMA A-configuration uniform weighting map (contours; synthesized beam=0$\farcs$41$\times$0$\farcs$30, P.A.=$-$49$^\circ$) over the 3.6 cm VLA natural weighting map (colors; synthesized beam=0$\farcs$32$\times$0$\farcs$24, P.A.=$-$14$^\circ$). The VLA map was made after concatenation of data from all the observed epochs. The rms of the VLA map is 19~$\mu$Jy~beam$^{-1}$. Contours are $-$3, 3, 6, 12, 24, and 48 times the rms of the 3~mm map, 460 $\mu$Jy~beam$^{-1}$. \textbf{(b)} VLA 1.3 cm map of IRS 3 (Torrelles et al. 1998). The rms of the map is 500~$\mu$Jy~beam$^{-1}$. The open ellipse in the right panel represents an orbit of 30~AU, from which we propose that the water maser spots are arising (see section \ref{SectionIRS3}). Water masers spots (Torrelles et al. 1998) are marked with red (V$_{LSR}$ $>$ 11.8 km s$^{-1}$), grey (7.8 km s$^{-1}$ $<$ V$_{LSR}$ $<$ 11.8 km s$^{-1}$) and blue (V$_{LSR}$ $<$ 7.8~km~s$^{-1}$) circles. In both panels, the (0,0) position corresponds to $\alpha$(J2000)=05$^h$47$^m$04.638$^s$, $\delta$(J2000)=00$^\circ$21$\arcmin$47.94$\arcsec$.}

\label{Fig7}  
\end{center}
\end{figure}

 In Figure \ref{Fig7} we show our 3.6~cm map of IRS~3 made after concatenation of data from all the epochs (Figure \ref{Fig7}a), as well as the 1.3 cm map (Figure \ref{Fig7}b). As can be seen in the figure, the cm emission at both wavelengths seems to be tracing the same structure, i.e., the radio jet. In Figure \ref{Fig7}a, we have also superposed the 3~mm contour map obtained from our CARMA A-configuration observations. The 3~mm source appears elongated, nearly perpendicular to the jet traced by the centimeter emission. From a Gaussian fit to the 3~mm source, we obtain a deconvolved size (FWHM) of 0$\farcs$32$\times$0$\farcs$14~$\pm$~0$\farcs$01 ($\sim$120$\times$50 AU), and a P.A. of 127$^\circ$~$\pm$~2$^\circ$, which is the same P.A. obtained by Torrelles et al. (1998) from the water maser spots. This strongly suggests that the 3 mm emission is tracing, at larger scales, the same structure than the water maser spots, i.e., a circumstellar disk. A similar situation has been found in G192.16 (Shepherd \& Kurtz 1999) and IRAS~16547$-$4247 (Franco-Hern\'andez et al. 2009), where the water masers trace a smaller structure than the dust.
 
\begin{figure}[!h]
\begin{center}

\includegraphics[width=\textwidth]{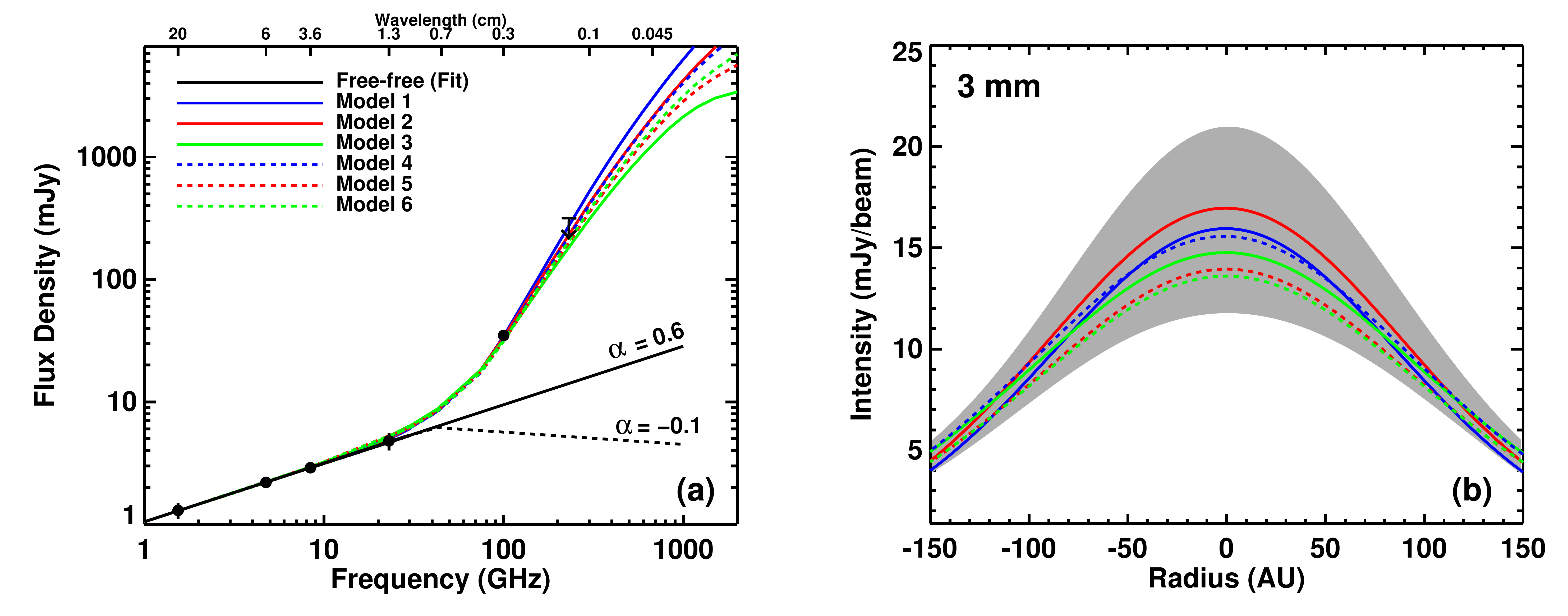}  

\caption{\textbf{(a)} Spectral energy distribution of IRS~3. Black circles are the total flux density observed at each wavelength. The upper limit at 1.3~mm is taken from previous lower angular resolution observations by Cortes et al. (2006). Solid black line is a least squares fit to the centimeter data. Dashed black line represents a lower limit to the contribution of the free-free emission at high frequencies, assuming that the jet is optically thin at wavelengths shorter than 1~cm. Color lines are the sum of the extrapolated free-free emission and the thermal dust emission obtained from disk models corresponding to stellar masses between 3 and 6~M$_\sun$ (see Table \ref{Tab4}). \textbf{(b)} Predicted intensity profiles at 3 mm along the major axis of the disk for the different disk models (color lines). The observed intensity profile (after correction from the estimated free-free contamination; see text) is represented by the grey area. The upper limit of the intensity profile is obtained assuming that the free-free emission is optically thin after 1~cm, while the lower limit is obtained assuming the same slope as in the 20-1~cm range (see panel a).}

\label{Fig8}  
\end{center}
\end{figure}

 In Figure \ref{Fig8} we show the SED of IRS~3 in the range from 20~cm to 1~mm. From a fit to the data in the range from 20 cm to 1.3 cm (solid line in Figure \ref{Fig8}a) we obtain a spectral index of 0.6~$\pm$~0.1, which is typical of radio jets expanding with constant velocity and constant opening angle (Reynolds 1986). As can be seen in this figure, the free-free emission (dashed line) can account for less than 30\% ($\sim$10 mJy) of the observed flux density at 3~mm (35~$\pm$~1~mJy). Therefore, the 3~mm emission is dominated by thermal dust emission, likely from a circumstellar disk. A crude estimate of the disk mass can be obtained from the 3~mm emission. Adopting a 3~mm dust opacity per gram of dust+gas of 0.006 cm$^2$ g$^{-1}$ (D'Alessio et al. 2001; assumes a gas-to-dust ratio of 100) and assuming an average temperature of 100 K, would result in a disk mass of $\sim$0.1 $M_\sun$. In order to test the plausibility of the disk interpretation using a more realistic calculation, we compared the observed millimeter flux densities and the 3 mm intensity profile along the major axis of IRS~3 with the predictions of an irradiated accretion disk model.

 The model calculation is described in detail in D'Alessio et al. (1998, 2001). In summary, it assumes an accretion disk with a density given by the conservation of angular momentum flux and a uniform mass accretion rate. We assume that the disk is heated by viscous dissipation (which is important close to the star and close to the midplane) and stellar irradiation (which is important in most of the disk structure). The energy is transported mainly by radiation, but convection and turbulent transfer are also included. The viscosity is calculated using the Shakura \& Sunyaev (1973) prescription, assuming a standard value of the viscosity parameter $\alpha_{t}=0.01$. The disk is assumed to have azimuthal symmetry, to be in hydrostatic equilibrium in the vertical direction, and in centrifugal balance with the stellar gravitational field in the radial direction. The disk self-gravity is assumed to be negligible compared to the stellar gravity.

 The model also accounts for dust evolution given by grain growth and settling towards the disk midplane. We parametrize two populations of dust, with different size distributions and dust-to-gas mass ratios, one corresponding to small depleted grains at the disk upper layers, and the other corresponding to big grains settled closer to the disk midplane (D'Alessio et al. 2006). For the population of small grains in the upper layer, we assume that their maximum size is  $a_{max}$~=~0.25~$\mu$m, which is a typical value for interstellar dust grains. For the population of big grains settled close to the midplane we assume a maximum grain size $a_{max}$~=~1~mm (Beckwith \& Sargent 1990; D'Alessio et al. 2001). For the grain composition, we assume a mixture of graphite and silicate compounds with abundances taken from Draine \& Lee (1984). 

 A grid of models was calculated for stars with masses in the range $M_*$ = 0.5-5~$M_\sun$ (stars with a higher mass would exceed the observed total luminosity of the region), disk mass accretion rates in the range $\dot M$ = 10$^{-8}$-10$^{-6}~M_\sun$~yr$^{-1}$ and disk radii in the range $R_d$ = 100-200~AU.  The evolutionary status of IRS~3 is not well stablished (Aspin et al. 1992; Walther et al. 1991; Tamura et al. 2007); thus, we have explored models assuming that the central star is either a protostar (that has accreted only a fraction of its final mass) or a pre-main-sequence (PMS) star near its birthline (that has almost reached its final mass). The values of the stellar luminosity, radius and effective temperature were taken from the evolutionary tracks for accreting protostars given by Wuchterl \& Tscharnuter (2003) and for PMS stars given by Siess et al. (2000) (in this last case, we have taken into account the age corrections in the determination of the birthline discussed by Hartmann 2003).

 The fixed parameters of the model are the disk inclination angle (angle between the polar axis and the line of sight), inferred from the aspect ratio of the 3~mm image to be $i=$60$^\circ$, the viscosity parameter $\alpha_{t}=0.01$, the maximum grain size of the small and big grain populations (0.25~$\mu$m and 1~mm respectively), and the degree of settling ($\epsilon$=0.1).
  
 The observational constraints are the flux density at 3~mm and the intensity profile along the major axis of the disk. We considered the flux density and the intensity profile of the uniform weighting map at 3~mm (which has the highest angular resolution) to avoid contamination from a possible circumstellar envelope. To take into account possible contamination from free-free emission of the jet, we obtained lower and upper limits to the observed flux density and intensity profile of the disk at 3~mm by subtracting Gaussian sources with the size of the beam (which represent the jet at 3~mm). The lower limit is obtained by considering that the flux density of the free-free emission at 3~mm is given by the extrapolation of the fit to the 20-1.3 cm data ($\alpha$=0.6), while, for the upper limit, we assumed that the free-free emission is optically thin ($\alpha$=$-$0.1) at wavelengths shorter than 1~cm (see Figure \ref{Fig8}). An additional constraint to the SED is
obtained from lower angular resolution (beam = 3$\farcs$7) observations at 1.3~mm by Cortes et al. (2006). Since this emission peaks at the position of the radio source IRS~3, we consider the value of the intensity peak ($\sim$300~mJy~beam$^{-1}$) as an upper limit to the flux density of the IRS~3 disk at 1.3~mm.
 
 We find that it is indeed possible to explain the observations using accretion disk models with typical parameters (see Table \ref{Tab4}). If we adopt for the central star the parameters of a PMS near its birthline (Siess et al. 2000; Hartmann 2003), we find that, regardless of the assumed mass accretion rate, disk models with the mass of the central star $M_*<3~M_\odot$ are not hot enough to produce the observed flux density at 3~mm. Thus, we constrain the central mass to be $3~M_\sun \leq M_* \leq 5~M_\sun$. Finally, among the PMS models with the appropriate central stellar mass, we find that only those with $\dot M$~=~2-5~$\times$~10$^{-7}~M_\sun$~yr$^{-1}$ can explain both the SED and the intensity profile at 3~mm. For these models, the disks have masses in the range $M_d$~=~0.5-1~$M_\sun$ (Table \ref{Tab4}). On the other hand, if we adopt a central protostar with an age of a few $10^5$ yr with the parameters taken from the models of Wuchterl \& Tscharnuter (2003), we find that the observational constraints can be reproduced for protostars whose accreted mass lies roughly in the range M$_*$=1.3-3 $M_\sun$ (corresponding to a final mass in the range 2-5 $M_\sun$, respectively). Central sources corresponding to less massive protostars require disks too massive that would be unstable while more massive protostars would exceed the luminosity constraints. The mass of the disk is in range M$_d$=0.5-0.7 $M_\sun$ and the mass accretion rate is in the range $\dot{M}$=3-5~$\times$~10$^{-7}~M_\sun$~yr$^{-1}$. In Table \ref{Tab4} we give a sample of valid disk models. Therefore, we conclude that although the currently available data are insufficient to tightly constrain the disk model, they are consistent with the presence of an accretion protoplanetary disk of 0.5-1 $M_\sun$ surrounding either an accreting protostar or a young PMS star of intermediate mass. Unfortunately, the evolutionary tracks by Siess et al. (2000) are uncertain near the birthline. Those presented by Wuchterl \& Tscharnuter (2003) are poorly sampled for intermediate-mass protostars and the stellar parameters we have adopted are only crude estimates obtained by interpolation. More realistic and detailed evolutionary calculations are requiered to infer better estimates of the disk parameters. Additional multiwavelength observations with an angular resolution high enough to isolate the emission of IRS 3 are required in order to complete the SED of this object. This will allow to constrain the bolometric luminosity and the evolutionary status of the source that are necessary to undertake a more detailed modeling.
 
 From the analysis of the spatio-kinematical distribution of water maser spots, Torrelles et al. (1998) proposed that these masers are orbiting a 1~$M_\sun$ star at an orbital radius of 20~AU. Our modeling of the disk suggests that the mass of the central star can be larger. The value of 1~$M_\sun$ was inferred by these authors by assuming that the water maser spots are tracing the full extent of an edge-on orbit of radius 20 AU. Since our 3~mm observations and modeling of the data suggest that the disk is seen with an inclination angle of 60$^\circ$, and extends over a larger region ($\sim$150~AU), it is possible that the masers could be tracing only a part of an orbit with a radius larger than 20 AU, impliying that the stellar mass derived by Torrelles et al. (1998) should be taken as a lower limit.

 In order to test if our results are compatible with the spatio-kinematical distribution of the water maser spots, we made the following analysis. The orbital velocity, $v$, at a given orbital radius, $r$, is given by
 
\begin{equation}
\left( \frac{v}{\rm km~s^{-1}} \right) = 29.8 \left( \frac{M_*}{\rm M_\sun} \right)^{1/2} \left( \frac{r}{\rm AU} \right)^{-1/2}, \label{eq1}
\end{equation}
 
\begin{figure}[!]
\begin{center}

\includegraphics[width=0.5\textwidth]{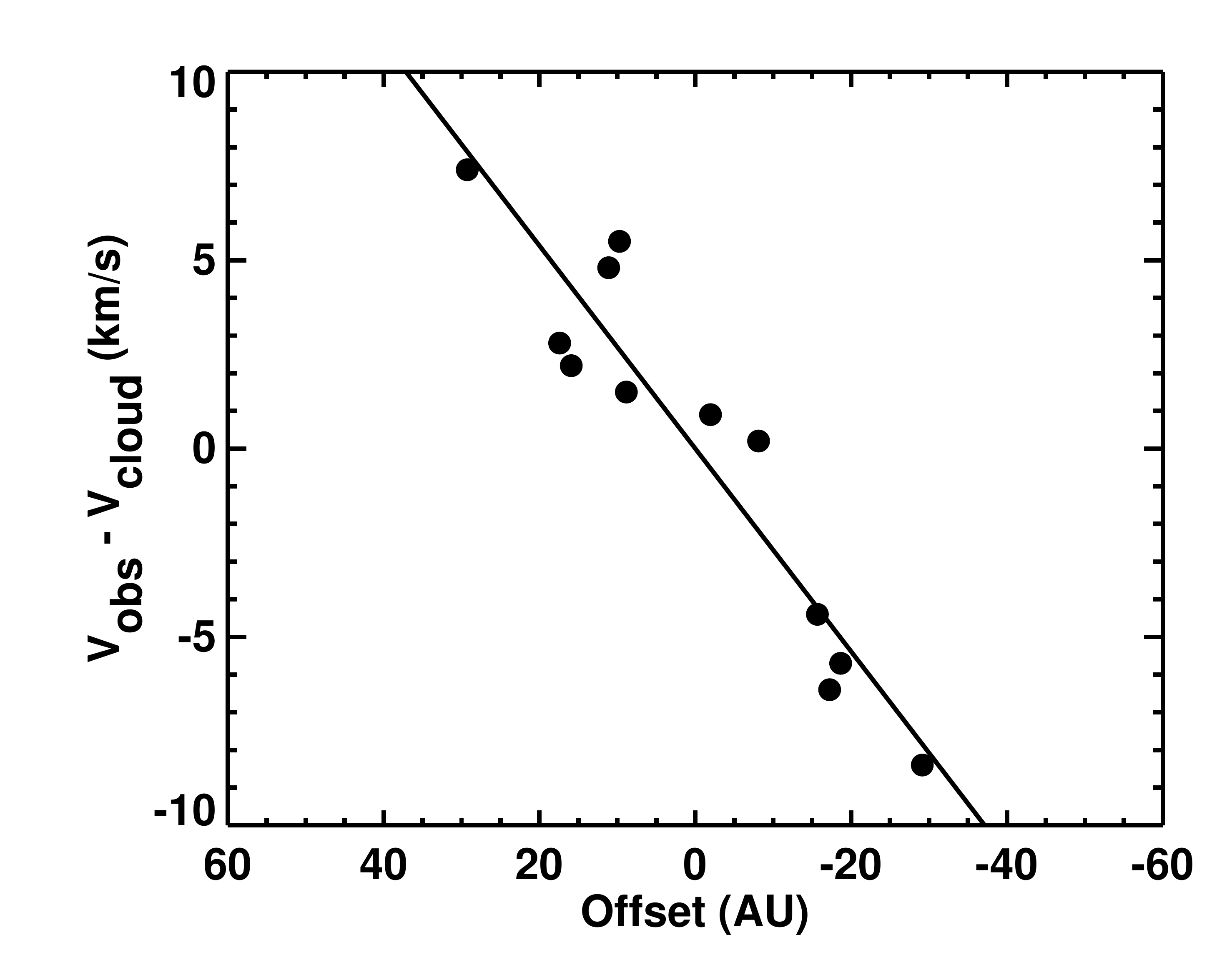}  

\caption{Position-velocity diagram of the water maser spots in IRS 3 observed by Torrelles et al. (1998). Velocities are relative to the systemic velocity of the cloud, assumed to be 9.8 km~s$^{-1}$ (Torrelles et al. 1998). Offsets are measured with respect to the maser with the velocity closest to the systemic velocity of the cloud. The solid line is a least squares fit to the data and indicates an observed gradient of 0.27~km~s$^{-1}$~AU$^{-1}$.}

\label{Fig9}  
\end{center}
\end{figure}

If all the masers are located at the same orbital radius $r$, then the observed velocity gradient of the masers will be given by 

\begin{equation}
\frac{\partial V_{\rm LSR}}{\partial l}=\frac{v}{r} \sin{i}, \label{eq2}
\end{equation}
 
\noindent where $i$ is the inclination angle of the disk. In Figure \ref{Fig9} we show a position-velocity (P-V) diagram of the water maser 
spots detected by Torrelles et al. (1998). From a least squares fit to this P-V diagram, we estimate a velocity gradient of $\partial V_{\rm 
LSR}/\partial l$~$\simeq$~0.27~km~s$^{-1}$~AU$^{-1}$. From the above equations, assuming an inclination angle $i\simeq$60$^\circ$ (estimated from the 3~mm continuum map), the observed velocity gradient, and the range of masses for the central star (predicted from the disk modeling), we obtain that the masers originate from an orbital radius with a value in the range 23-36 AU, depending on the mass of the central star. We illustrate this scenario in Figure \ref{Fig7}b where we show an ellipse that represents a 30~AU orbit in the disk.

\begin{figure}[!h]
\begin{center}

\includegraphics[height=0.75\textheight]{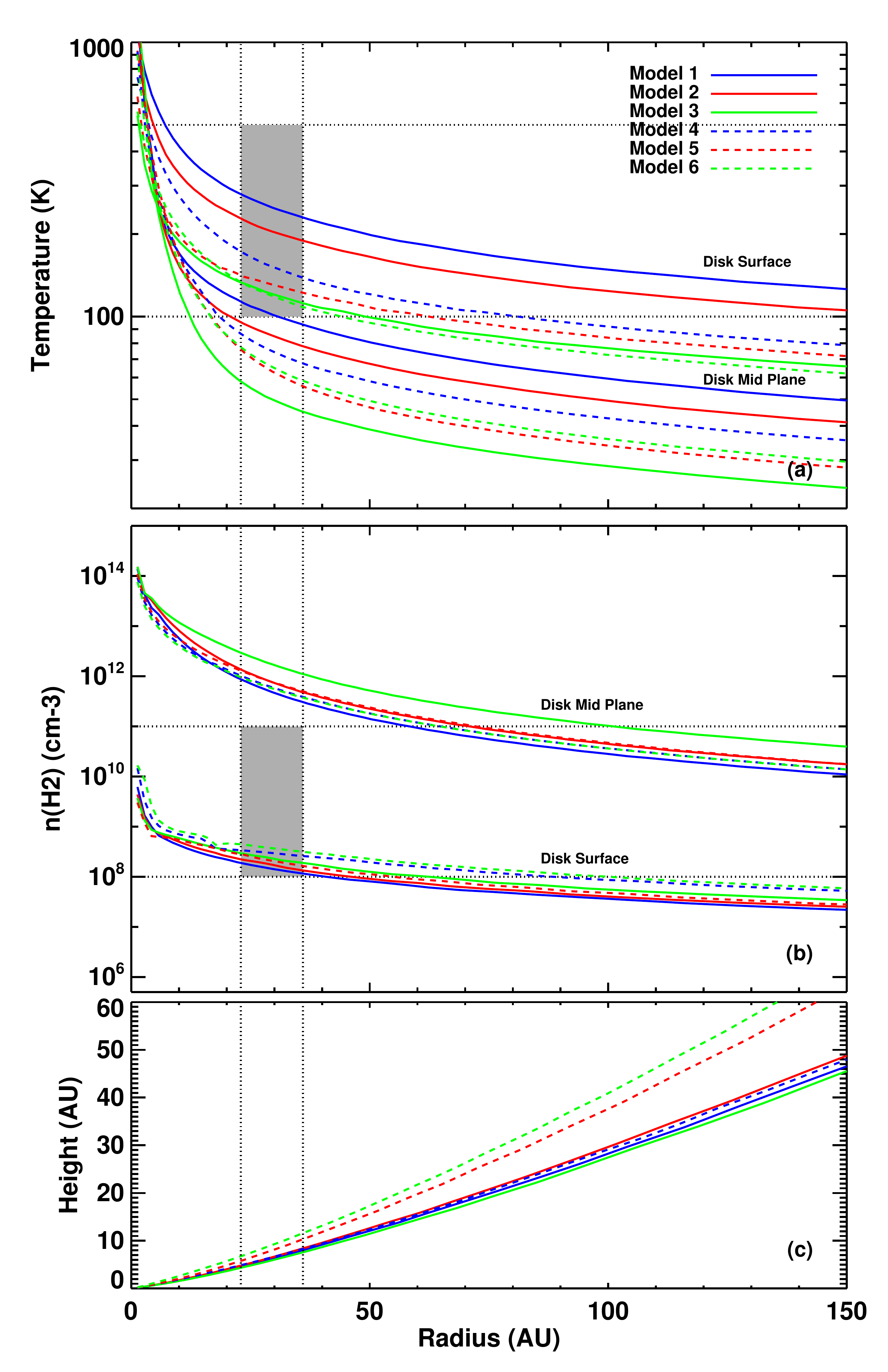}  

\caption{Temperature (panel a), density (panel b) and height (panel c) profiles, obtained for different models of the IRS~3 disk. In panels (a) and (b), the horizontal dotted lines mark the rough limit conditions of temperature and density for water maser emission (Elitzur et al. 1989). The vertical dotted lines mark orbits at radii of 23 and 36 AU, the proposed range of distances where the water maser emission detected by Torrelles et al. (1998) originates (see section \ref{SectionIRS3}). The grey area in each panel marks the region constrained by the disk radii where the temperature (or density) condition for the maser emission is fulfilled. As can be seen, all the models have proper density and temperature conditions near the disk surface to produce water maser emission in the radius range 23-36~AU.}

\label{Fig10}  
\end{center}
\end{figure}

 In Figure \ref{Fig10} we show the density and temperature profiles at the disk surface and at the disk mid-plane for each model. We also show in this figure (horizontal dotted lines) rough limits of physical conditions for water maser emission (Elitzur et al. 1989). As can be seen in the figure, all the models have appropriate density and temperature conditions near the disk surface to produce water maser emission in the radius range 23-36~AU. Therefore, the physical structure (temperature and density) of the disks resulting from our modeling seems to be consistent with the observed water maser emission previously found in the IRS~3 disk.
   
\section{Conclusions}

 We carried out observations toward the intermediate-mass star-forming region NGC~2071 with the VLA and CARMA. We detected cm and mm emission from IRS~1, IRS~2, IRS~3 and VLA~1. Our main conclusions can be summarized as follows:
 
\begin{itemize}

 \item All the sources show an SED consistent with free-free thermal emission at cm wavelengths and thermal dust emission at 3~mm, suggesting that all of them are young stellar objects. 
 
 \item We have resolved IRS~2 in two components separated by $\sim$500~AU, suggesting that IRS~2 is actually a young binary system. 

 \item Source IRS~1 shows a complex morphology at centimeter wavelengths. We discuss the possibility that the morphology of this source could be consequence of strong interactions of the jet with the ambient medium, or that we are detecting emission from a binary jet. Our highest angular resolution 1.3~cm map of this source suggests the presence of a binary system (separation between components $\simeq$40~AU), supporting the hypothesis of a binary jet as the origin of the morphology of IRS~1. 
 
 \item The 3.6~cm emission of IRS~3 shows a clear jet-like shape. We observed time-variations in the morphology and flux density of the source that we interpret as due to an ejection of material and precession of the jet axis. We estimated the velocity of this ejection to be $\sim$140~km~s$^{-1}$, which is similar to the typical velocities observed in jets from low- and intermediate-mass YSOs.
 
 \item The 3~mm emission associated with IRS 3 seems to be tracing thermal dust emission arising from an angularly resolved circumstellar disk. An irradiated accretion disk model with a mass of 0.5-1 $M_\sun$ and a radius of $\sim$ 150 AU around an intermediate-mass YSO can account for the observed SED and spatial intensity profile at 3~mm of IRS 3, supporting this interpretation.
 
 \item Our results suggest that the water maser spots previously detected in IRS~3 by Torrelles et al. (1998) arise from a region in the disk located in an orbit of radius 24-36~AU.

\end{itemize}

\emph{Acknowledgements.} We thank N. Calvet and an anonymous referee for valuable comments. C.C.-G. acknowledges the financial support by the European Research Council for the ERC Advanced Grant \emph{GLOSTAR} under contract no. 247078. G.A., C.C.-G., J.F.G., M.O., and J.M.T. acknowledge support from MICINN (Spain) grant AYA2008-06189-C03 (co-funded with FEDER funds). C.C.-G., G.A., J.F.G., and M.O. acknowledge partial support from Junta de Andaluc\'{\i}a (TIC-126). L.F.R. and P.D. acknowledge the support of DGAPA, UNAM, and CONACyT (M\'exico).



\begin{deluxetable}{ccccccccrc}
\tabletypesize{\scriptsize}
\tablewidth{0pt}
\tablecaption{Summary of Observations\tablenotemark{a} \label{Tab1}}
\startdata
\hline \hline 
	              &       &       &		    &	         &	      &        	  & \multicolumn{2}{c}{Synthesized Beam\tablenotemark{c}}    &     \\ \cline{8-9}
                      &       &	      & Observation & Flux       & Phase      &   S$_{\rm phase}$\tablenotemark{b}	  &	   HPBW        & \multicolumn{1}{c}{P.A.}  & rms Noise\tablenotemark{c} \\
 Wavelength           & Conf. & Epoch & Date	    & Calibrator & Calibrator &       (Jy)		  &	 (arcsec)      & \multicolumn{1}{c}{(deg)} & ($\mu$Jy beam$^{-1}$)	\\ \hline
  20 cm                  & A & 1983.9 & 83-Nov-11 & 3C48    & J0530$+$135 & 1.90 $\pm$ 0.03  & 1.44 $\times$ 1.07 &  $-$18   &  75  \\

   6 cm                  & A & 1981.1 & 81-Feb-08 & 3C286   & J0530$+$135 & 4.41 $\pm$ 0.039 & 0.71 $\times$ 0.50 &  $-$39   &  40  \\

  3.6 cm                 & A & 1995.6 & 95-Aug-12 & 3C286   & J0541$-$056 & 1.225 $\pm$ 0.003  & 0.34 $\times$ 0.25 &  $-$15 &  27  \\
  3.6 cm                 & A & 1997.0 & 97-Jan-10 & 3C48    & J0541$-$056 & 0.809 $\pm$ 0.005  & 0.31 $\times$ 0.25 &	7 &  19  \\
  3.6 cm                 & A & 1998.2 & 98-Mar-27 & 3C48    & J0541$-$056 & 0.720 $\pm$ 0.004  & 0.35 $\times$ 0.25 &  $-$20 &  18  \\
  3.6 cm                 & A & 1998.4 & 98-May-26 & 3C48    & J0541$-$056 & 0.858 $\pm$ 0.008  & 0.36 $\times$ 0.24 &  $-$26 &  20  \\
  3.6 cm                 & A & 1999.5 & 99-Jul-03 & 3C48    & J0541$-$056 & 1.021 $\pm$ 0.005  & 0.34 $\times$ 0.27 &  $-$12 &  19  \\ 

   2 cm                  & A& 1987.5 & 87-Jul-04  & 3C286   & J0541$-$056 & 1.09 $\pm$ 0.01  & 0.18 $\times$ 0.15 &  $-$12   &  60  \\

  3 mm\tablenotemark{d}  & B & 2008.0 & 07-Dec-22 &  Uranus & J0541$-$056 & 1.06 $\pm$ 0.03  & 0.52 $\times$ 0.48 &     86   & 370  \\
  3 mm\tablenotemark{d}  & A & 2009.0 & 09-Jan-30 &  Uranus & J0541$-$056 & 0.53 $\pm$ 0.02  & 0.48 $\times$ 0.31 &    -21   & 250  \\ 
\enddata 

\tablenotetext{a}{Observations carried out with the VLA, except for the 3~mm observations that were carried out with CARMA.}

\tablenotetext{b}{Bootstrapped flux density of the phase calibrator.}

\tablenotetext{c}{For naturally weighted maps.}

\tablenotetext{d}{Bandpass calibration was achieved by observing J0530+135.}

\end{deluxetable}


\begin{deluxetable}{ccccccc}
\tablewidth{0pt}
\tablecaption{Spectral Energy Distributions of the Sources\tablenotemark{a}\label{Tab2}}
\startdata
\hline \hline 
           &      &			 \multicolumn{5}{c}{Flux Density (mJy)} 		 \\ \cline{3-7} 
Wavelength & Conf.&	IRS 1	  &	  IRS 2A     &     IRS 2B      &      IRS 3	&  VLA 1 \\ \hline
20 cm      &  A   &  4.0 $\pm$ 0.2 &	  $<$0.5     &      $<$0.5     &  1.3 $\pm$ 0.2 &	$<$0.5      \\
6 cm       &  A   &  6.4 $\pm$ 0.3 &  0.5  $\pm$ 0.1    & 0.5  $\pm$ 0.1  &  2.2 $\pm$ 0.2 &  0.34 $\pm$ 0.08  \\
3.6 cm     &  A   &  7.7 $\pm$ 0.9\tablenotemark{b} &  0.42 $\pm$ 0.05\tablenotemark{b}   & 0.25 $\pm$ 0.06\tablenotemark{b} &  2.9 $\pm$ 0.2\tablenotemark{b} &  0.34 $\pm$ 0.07\tablenotemark{b}  \\
2 cm       &  A   & 11.0 $\pm$ 3   &	  $<$0.4     &      $<$0.4     &      $>$2.2	&  0.5  $\pm$ 0.2   \\
1.3 cm     &  A   & 14.0 $\pm$ 2.0\tablenotemark{c} &	  $<$1.0\tablenotemark{c}     &      $<$1.0\tablenotemark{c}     &  4.8 $\pm$0.8\tablenotemark{c}  &  0.9  $\pm$ 0.3\tablenotemark{c}   \\
3 mm       &  B   & 36   $\pm$ 1   &  5    $\pm$ 1      & 5    $\pm$ 1    & 42   $\pm$ 1	&  9	$\pm$ 1     \\ 
3 mm       &  A   & 23   $\pm$ 1   &       $<$2	     &       $<$2      & 35   $\pm$ 1   &  6	$\pm$ 1     \\ \hline
\enddata

\tablenotetext{a}{Flux densities obtained with the task IMSTAT of AIPS.}
\tablenotetext{b}{Average of the values obtained at each epoch. Quoted uncertainties are the dispersion of the flux density over all the epochs.}
\tablenotetext{c}{From Trinidad et al. (2009).}
\end{deluxetable}


\begin{deluxetable}{ccccc}
\tablewidth{0pt}
\tablecaption{Monitoring of IRS 3 at 3.6 cm\tablenotemark{a} \label{Tab3}}
\startdata
\hline \hline 
           &  Major Axis     & Position Angle & Peak Intensity  & Flux Density  \\
 Epoch     &  (arcseconds)   &    (degrees)   &  (mJy/beam)     &     (mJy)     \\ \hline
1995.6     & 0.44 $\pm$ 0.02 &  11 $\pm$ 1    & 1.29 $\pm$ 0.05 & 3.0 $\pm$ 0.2 \\
1997.0     & 0.43 $\pm$ 0.02 &  12 $\pm$ 1    & 1.11 $\pm$ 0.05 & 2.5 $\pm$ 0.1 \\
1998.2     & 0.42 $\pm$ 0.02 &  12 $\pm$ 1    & 1.23 $\pm$ 0.04 & 2.6 $\pm$ 0.1 \\
1998.4     & 0.38 $\pm$ 0.02 &  15 $\pm$ 1    & 1.53 $\pm$ 0.04 & 3.0 $\pm$ 0.1 \\
1999.5     & 0.55 $\pm$ 0.02 &  16 $\pm$ 1    & 1.06 $\pm$ 0.05 & 2.8 $\pm$ 0.2 \\ \hline
\enddata
\tablenotetext{a}{Parameters obtained from Gaussian fits to the IRS~3 source in the uniform weighting maps.}
\end{deluxetable}


\begin{deluxetable}{ccccccc}
\tablewidth{0pt}
\tablecaption{Parameters of Selected Disk Models \label{Tab4}}
\startdata
\hline \hline 

 Disk  & $M_d$\tablenotemark{a}  & $R_d$\tablenotemark{b} & $\dot M$\tablenotemark{c} & $M_*$\tablenotemark{d}  & $M_f$\tablenotemark{e} &  Stellar \\
Model &(M$_\sun$) & (AU)  & (M$_\sun$ yr$^{-1}$) & (M$_\sun$)  & (M$_\sun$)  &  Model\tablenotemark{f} \\ 
\hline
 M1   &  0.8      &  175                     & 2$\times$10$^{-7}$ &	 5	 &  $\sim$5     &	PMS        \\
 M2   &  0.5      &  150                     & 3$\times$10$^{-7}$ &	 4	 &  $\sim$4     &	PMS        \\
 M3   &  1        &  175                     & 4$\times$10$^{-7}$ &	 3	 &  $\sim$3     &	PMS        \\

 M4   &  0.5      &  175                     & 3$\times$10$^{-7}$ &	 3	 &  $\sim$5     &	 Protostar  \\
 M5   &  0.6      &  175                     & 4$\times$10$^{-7}$ &	 1.8	 &  $\sim$2.5   & Protostar  \\
 M6   &  0.7      &  175                     & 5$\times$10$^{-7}$ &	 1.3	 &  $\sim$2     &	Protostar  \\
\enddata
\tablenotetext{a}{Mass of the disk.}
\tablenotetext{b}{Disk outer radius. Note that the diameter of the disk is significantly larger than the deconvolved FWHM obtained from a Gaussian fit.}
\tablenotetext{c}{Mass accretion rate.}
\tablenotetext{d}{Accreted mass of the central star.}
\tablenotetext{e}{Final mass of the central star at the end of the accretion process.}
\tablenotetext{f}{PMS = Parameters of the central star taken from the PMS evolutionary tracks of Siess et al. (2000). Protostar = Parameters of the central star taken from the protostellar evolutionary tracks of Wuchterl \& Tscharnuter (2003).}
\end{deluxetable}



\end{document}